\documentclass{elsart}

\newcounter{bla}
\newenvironment{refnummer}{%
\list{[\arabic{bla}]}%
{\usecounter{bla}%
 \setlength{\itemindent}{0pt}%
 \setlength{\topsep}{0pt}%
 \setlength{\itemsep}{0pt}%
 \setlength{\labelsep}{2pt}%
 \setlength{\listparindent}{0pt}%
 \settowidth{\labelwidth}{[9]}%
 \setlength{\leftmargin}{\labelwidth}%
 \addtolength{\leftmargin}{\labelsep}%
 \setlength{\rightmargin}{0pt}}}
 {\endlist}
 
\usepackage{bm}
\usepackage{dcolumn}
\usepackage{graphicx}
\usepackage{color}
\usepackage{dirtree}
\usepackage{amssymb}
\usepackage{verbatim}
\usepackage{listings}

\newcommand{\code}[1]{{\ttfamily #1}}

\begin{document}
\begin{frontmatter}


\title{Nonlinear Boltzmann equation for\\
the homogeneous isotropic case:\\
Minimal deterministic Matlab program}

\author{Pietro Asinari}
\address{Department of Energetics, Politecnico di Torino,\\
Corso Duca degli Abruzzi 24, Torino, Italy\\
email: pietro.asinari@polito.it, web: http://staff.polito.it/pietro.asinari}

\begin{abstract}
The homogeneous isotropic Boltzmann equation (HIBE) is a fundamental dynamic model for many applications in thermodynamics, econophysics and sociodynamics. Despite recent hardware improvements, the solution of the Boltzmann equation remains extremely challenging from the computational point of view, in particular by deterministic methods (free of stochastic noise). This work aims to improve a deterministic direct method recently proposed [V.V. Aristov, Kluwer Academic Publishers, 2001] for solving the HIBE with a generic collisional kernel and, in particular, for taking care of the late dynamics of the relaxation towards the equilibrium. Essentially (a) the original problem is reformulated in terms of particle kinetic energy (exact particle number and energy conservation during microscopic collisions) and (b) the computation of the relaxation rates is improved by the DVM-like correction, where DVM stands for Discrete Velocity Model (ensuring that the macroscopic conservation laws are exactly satisfied). Both these corrections make possible to derive very accurate reference solutions for this test case. Moreover this work aims to distribute an open-source program (called \code{HOMISBOLTZ}), which can be redistributed and/or modified for dealing with different applications, under the terms of the GNU General Public License. The program has been purposely designed in order to be minimal, not only with regards to the reduced number of lines (less than 1,000), but also with regards to the coding style (as simple as possible). 


\end{abstract}

\end{frontmatter}


{\bf PROGRAM SUMMARY}

\begin{small}
\noindent
{\em Manuscript Title:} Nonlinear Boltzmann equation for the homogeneous isotropic case: Minimal deterministic Matlab program \\
{\em Authors:} Pietro Asinari \\
{\em Program Title:} \code{HOMISBOLTZ} \\
{\em Journal Reference:}                                      \\
{\em Catalogue identifier:}                                   \\
{\em Licensing provisions:} 
  The program is free software, which can be redistributed and/or modified under the terms of the 
  GNU General Public License. \\
{\em Programming language:} Tested with Matlab\textregistered$\;$ version $\geq$ 6.5. However, in principle, any recent version of Matlab\textregistered$\;$ or Octave should work. \\
{\em Computer:} All supporting Matlab\textregistered$\;$ or Octave \\
{\em Operating system:} All supporting Matlab\textregistered$\;$ or Octave \\
{\em RAM:} 300 MBytes \\
{\em Number of processors used:}                              \\
{\em Supplementary material:}                                 
{\em Classification:} 23 Statistical Physics and Thermodynamics \\
{\em External routines/libraries:}                             \\
{\em Subprograms used:}                                       \\
{\em Nature of problem:}\\
   The problem consists in integrating the homogeneous Boltzmann equation for a generic collisional 
   kernel in case of isotropic symmetry, by a deterministic direct method. Difficulties arise from 
   the multi-dimensionality of the collisional operator and from satisfying the conservation of particle number 
   and energy (momentum is trivial for this test case) as accurately as possible, in order to preserve 
   the late dynamics. \\
{\em Solution method:}\\
   The solution is based on the method proposed by Aristov [1], but with two substantial improvements: 
   (a) the original problem is reformulated in terms of particle kinetic energy (this allows one to ensure exact 
   particle number and energy conservation during microscopic collisions) and (b) a DVM-like correction 
   (where DVM stands for Discrete Velocity Model) is adopted for improving the relaxation rates (this 
   allows one to satisfy exactly the conservation laws at macroscopic level, which is particularly important for 
   describing the late dynamics in the relaxation towards the equilibrium). Both these corrections make possible to 
   derive very accurate reference solutions for this test case. \\
{\em Restrictions:}\\
   The nonlinear Boltzmann equation is extremely challenging from the computational point of view, in particular 
   for deterministic methods, despite the increased computational power of recent hardware. In this work, only the 
   homogeneous isotropic case is considered, for making possible the development of a minimal program (by a simple 
   scripting language) and allowing the user to check the advantages of the proposed improvements beyond the 
   Aristov's method [1]. The initial conditions are supposed parameterized according to a fixed analytical 
   expression, but this can be easily modified. \\
{\em Unusual features:}\\
   There are no unusual features. \\
{\em Additional comments:}\\
  There are no additional comments. \\
{\em Running time:}\\
  From minutes to hours (depending on the adopted discretization of the kinetic energy space). For example, on a 
  64 bit workstation with Intel\textregistered$\;$ Core\texttrademark$\;$ i7-820Q Quad Core CPU at 1.73 GHz and 8 
  MBytes of RAM, the provided test run (with the corresponding binary data file storing the pre-computed relaxation 
  rates) requires 154 seconds.\\
{\em References:}
\begin{refnummer}
\item V.V. Aristov, {\em Direct Methods for Solving the Boltzmann Equation and Study of Nonequilibrium Flows}, 
	Kluwer Academic Publishers, 2001.
\end{refnummer}
\end{small}


\section{Introduction}

In a dilute gas, the Boltzmann transport equation \cite{boltzmann1872, cercignani1987} describes the time evolution of the single-particle distribution function, which provides a statistical description about the positions and velocities of the gas molecules. From the theoretical point of view, it is one of the most important equations of non-equilibrium statistical mechanics and one the most powerful paradigms for explaining transport phenomena in fluids. Moreover, from the engineering point of view, since early fifties, it received a lot of attention due to aerodynamic requirements for high altitude vehicles and vacuum technology requirements \cite{cercignani1987}. Nowadays, the set of applications has been widen by including dilute gas flows in micro-electro-mechanical systems (MEMs) \cite{cercignani2006}. These devices are increasingly applied to a great variety of industrial and medical problems. In these problems, given the small dimensions of the devices, it is necessary to use the kinetic theory, instead of the usual fluid dynamics, based on the Navier-Stokes equations, to describe the motion of dilute gases in the small gaps of these devices.

Because of the intrinsic complexity of this equation (the single-particle distribution function is defined in the phase space and the time evolution is ruled by a five-fold collisional integral), solving the nonlinear Boltzmann equation is extremely complex. Hence, from the very beginning, there was an attempt to formulate simpler models, which preserve the main features of the dynamic approach to the thermodynamic equilibrium. As pointed out in the Cercignani's biographical work \cite{cercignani1998}, Boltzmann himself started in his fundamental paper \cite{boltzmann1872} by considering first the case when the distribution function does not depend on space ({\em homogeneous} case), but only on time and the magnitude of the molecular velocity ({\em isotropic} collisional integral). The same homogeneous isotropic case is considered by Truesdell \cite{truesdell1966} in his famous lectures on natural philosophy, as the starting point for investigating the role of time in classical thermodynamic systems (which are assumed homogeneous by definition). In fact, despite the isotropy of the collisional integral, the actual time evolutions of the distribution function (far from the equilibrium) may be very different, depending on the initial conditions.

Concerning gas dynamics, focusing on the homogeneous isotropic case, it may seem a bit limiting. For example, an immediate consequence of the isotropic symmetry is that all the odd statistical moments are null by definition and hence (meaningful) moment equations can be derived for even moments only. However it is well known that the decomposition between even moments (pressure, energy,...) and odd moments (momentum, thermal flux,...) is a key concept in deriving the fluid dynamic description from the full Boltzmann equation, in case of vanishing Knudsen number \cite{sone2002}. In particular, in recovering the incompressible limit of the Navier-Stokes equations, the Mach number is assumed as small as the Knudsen number (diffusive scaling, see \cite{sone2002}) and hence the kinetic description collapses in a small neighborhood of the statistical core defined by the even moments only. This means that the distribution function can be expanded around an equilibrium distribution function, which depends on the even moments only. Hence describing properly the manifold defined by the even moments is the first basic step for describing the dynamics due to small deviations from the local equilibrium. This is the key idea behind the derivation of the so-called Lattice Boltzmann Method (LBM) \cite{succi2001}. This is the reason why the even moments are sometimes called backbone moments of the LBM description \cite{karlin2010}. A similar idea holds for the so-called quadrature method of moments (QMOM), which is a generic solution method for population balance models \cite{marchisio2005}. The common feature between LBM and QMOM is that both solve moment systems of equations, which are based on a contraction of the statistical description given by the Boltzmann equation. The systematic derivation of moment equations from the Boltzmann equation is beyond the purposes of the present work: a detailed review can be found in Ref. \cite{struchtrup2005}.

The interest with regards to the homogeneous isotropic Boltzmann equation goes beyond simple dilute gases. In the so-called econophysics \cite{chatterjee2005}, a Boltzmann type model is sometimes introduced for studying the distribution of wealth in a simple market. The founding idea, dating back to the works of Mandelbrot \cite{mandelbrot1960}, is that the laws of statistical mechanics govern the behavior of a huge number of interacting individuals just as well as that of colliding particles in a gas container. The classical theory for homogeneous gases is easily adapted to the new economic framework: molecules and their velocities are replaced by agents and their wealth, and instead of binary collisions, one considers trades between two individuals. The goal is to recover the macroscopic distributions of wealth by tuning the microscopic models for the binary interaction among the agents. The parameters of the microscopic model can be either constant or random quantities. A recent review on this topic can be found in Ref. \cite{during2008} and the references therein.

Another recent application of the homogeneous isotropic Boltzmann equation is given by opinion formation modeling in quantitative sociology, also called sociodynamics or sociophysics \cite{weidlich2000}. Quantitative sociology has the ambitious aim to provide a general strategy, that means a frame of theoretical concepts, for designing mathematical models for the quantitative description of a rather broad class of collective dynamical phenomena within human society, in particular opinion formation. The modeling of opinion dynamics has been treated in numerous works, because of its application to politics, to predict the behavior of voters during an election process or the public opinion tendencies \cite{weidlich2000}. Classical kinetic models based on homogeneous isotropic Boltzmann-like equations can be derived by prescribing the collision kernel for the microscopic particle interactions, namely the sociophysical model which prescribes the exchange rules for opinion in a binary interaction \cite{helbing1995}.

Since the Boltzmann equation was the starting point for constructing numerous kinetic equations in many fields of physics, many numerical techniques have been proposed to solve it. A complete review of these efforts is clearly beyond the purposes of the present work: however a complete discussion can be found in the Ref. \cite{aristov2001} and the references therein. Despite this wide scenario of numerical methods and the constant increase in the computational power, solving the Boltzmann equation in practical applications is still challenging nowadays. In particular, the most demanding step consists in the evaluation of the collisional integral (which is in general five-fold in three dimensions). It is possible to distinguish between (a) {\em stochastic} and (b) {\em deterministic} methods in evaluating the collisional integral. In the stochastic methods, like the Monte Carlo method, one uses a combination of approximations based on randomly-generated variables and a fixed (molecular) velocity grid. On the other hand, in deterministic (or direct) methods, one uses only regular lattices in velocity space, usually dealing with a larger computational effort in order to achieve better accuracy (free of stochastic noise) \cite{aristov2001}. 

The goal of this work is twofold.
\begin{itemize}
\item First of all, this work aims to improve the deterministic numerical method proposed by Aristov \cite{aristov2001} by (i) reformulating the original problem in terms of particle kinetic energy (this allows one to ensure exact particle number and energy conservation; momentum is trivially conserved because of the isotropic symmetry) and (ii) improving the computation of the relaxation rates (making it particularly suitable for dealing with the late dynamics of the relaxation towards the equilibrium).
\item Secondly, this work aims to distribute an open-source program (as minimal as possible) for solving by a deterministic method the homogeneous isotropic Boltzmann equation, which can be easily understood and modified for dealing with different applications (thermodynamics, econophysics and sociodynamics), in order to derive reliable reference solutions (with an accuracy which can not be easily obtained by stochastic methods). 
\end{itemize}

The paper is organized as follows. First, in Section \ref{background}, some theoretical background is provided about the homogeneous isotropic Boltzmann equation: in particular, the derivation of the homogeneous isotropic case, the energy formulation, the numerical method, the proposed correction for the relaxation rates and the adopted quadrature formulas are discussed. In Sections \ref{overview} and \ref{components}, an overview of the program structure and a description of the essential components are provided. Finally, in Sections \ref{installation} and \ref{test}, the instructions about installation and how to run a test case are provided.

\section{\label{background}Theoretical background}

\subsection{Boltzmann equation for Maxwell molecules}

Let us consider a dilute gas made of molecules. Let us introduce the probability density function, or distribution function, $f(t, \bm{x}, \bm{\xi})$ for the time $t\in\mathbb{R}^+$, for the position $\bm{x}\in\mathbb{R}^3$, where $\mathbb{R}^3$ is the physical space, and for the molecular velocity $\bm{\xi}\in\mathbb{R}_{\bm{\xi}}^3$, where $\mathbb{R}_{\bm{\xi}}^3$ is the velocity space ($\mathbb{R}^3\cup\mathbb{R}_{\bm{\xi}}^3$ is the phase space). Hence the distribution function is defined on the domain $\{t>0,\,\bm{x}\in\mathbb{R}^3,\,\bm{\xi}\in\mathbb{R}_{\bm{\xi}}^3\}$. The distribution function allows one to compute the infinitesimal probability to find some molecules in the time interval between $t$ and $t+dt$, in the infinitesimal volume $d\bm{x}\in\mathbb{R}$ around the point $\bm{x}$ and with a velocity in the infinitesimal volume $d\bm{\xi}\in\mathbb{R}_\xi$ around the velocity $\bm{\xi}$, namely $f(t, \bm{x}, \bm{\xi})dt d\bm{x} d\bm{\xi}$. According to the kinetic theory of gases, the probability density function of a dilute gas with elastic binary interactions satisfies the Boltzmann transport equation \cite{boltzmann1872,cercignani1987}, namely
\begin{equation}\label{BE}
\frac{\partial f}{\partial t}
+\bm{\xi}\cdot\nabla_{\bm{x}} f=Q(f,f),
\end{equation}
where the collisional integral is given by
\begin{equation}\label{coll}
Q(f,f)\dot{=}\int_{\mathbb{R}_{\bm{\xi}}^3}\int_{\bm{g}\cdot\bm{n}<0}
B(\bm{g},\bm{n})\left[f(\bm{\xi}') f(\bm{\xi_*}')-f(\bm{\xi}) f(\bm{\xi_*})\right]
d\bm{n}\,d\bm{\xi_*},
\end{equation}
$\bm{\xi_*}\in\mathbb{R}_{\bm{\xi}}^3$ is the generic field particle (integration dummy variable) and $d\bm{\xi_*}$ is its infinitesimal volume in the velocity space; $\bm{\xi}',\,\bm{\xi_*}'\in\mathbb{R}_{\bm{\xi}}^3$ are the post--collision test and field particle velocities respectively; $\bm{n}\in\mathbb{R}^3$ is the unit vector along the direction connecting the centers of the two particles during the instantaneous collision and versus pointing from particle $\bm{\xi}$ to $\bm{\xi}_*$, while $d\bm{n}$ is the infinitesimal solid angle; $\bm{g}=\bm{\xi}_*-\bm{\xi}$ is the relative velocity (of the field particle with regards to the test particle); finally, $B(\bm{g},\bm{n})$ is a volumetric particle flux or collision kernel. In the following, we will discuss only the case of Maxwell molecules \cite{cercignani1987}, which highly simplify the expression of the collision kernel, namely $B(|\bm{g}\cdot\bm{n}|)$. Let us assume the following expression
\begin{equation}\label{generic}
B(|\bm{g}\cdot\bm{n}|)=a^2\,c_s\,\left(\frac{\left|\,\bm{g}\cdot\bm{n}\,\right|}{c_s}\right)^\theta,
\end{equation}
where $a$ is the particle radius, $c_s$ is a characteristic mean particle velocity (i.e. statistical mean of the particle velocity deviations, which is related to the macroscopic sound speed) and $\theta$ is a tunable parameter (natural number, i.e. $\theta\in\mathbb{N}$). The case $\theta=1$ recovers the hard spheres model (popular in fluid dynamics), while the case $\theta=0$ recovers the constant kernel model (which yields constant collision frequency, as commonly done in econophysics and sociophysics). In the previous equation, the post--collision test and field particle velocities $\bm{\xi}'$ and $\bm{\xi_*}'$ are given by
\begin{eqnarray}
\label{post1}
\bm{\xi}'&=&\bm{\xi}+\left(\bm{g}\cdot\bm{n}\right)\,\bm{n},\\
\label{post2}
\bm{\xi_*}'&=&\bm{\xi_*}-\left(\bm{g}\cdot\bm{n}\right)\,\bm{n},
\end{eqnarray}
which means that there are many possible outcomes $(\bm{\xi}',\bm{\xi_*}')$ from a given pair of incoming (test and field) particle velocities $(\bm{\xi},\bm{\xi_*})$, depending on the impact direction $\bm{n}$ obtained by connecting the particle centers during the collision. With other words, the generic microscopic collision is defined once two additional degree of freedoms are specified ($\bm{n}$ is a versor).

\subsection{Homogeneous isotropic case}

Let us consider first the homogeneous case (in space). Consequently the probability density function becomes $f(t, \bm{\xi})$ and the homogeneous Boltzmann equation becomes
\begin{equation}\label{HBE}
\frac{\partial f}{\partial t}=Q(f,f),
\end{equation}
where the collisional integral is rewritten equivalently as
\begin{equation}\label{coll2}
Q(f,f)=\int_{-\infty}^{+\infty}\int_0^{4\pi}
{S}(q)\,B(q)\left[f(\bm{\xi}') f(\bm{\xi_*}')-f(\bm{\xi}) f(\bm{\xi_*})\right]
d\bm{n}\,d\bm{\xi_*},
\end{equation}
where $q=\bm{g}\cdot\bm{n}$, $B(q)=B(|q|)$ for simplicity and $S(q)$ is an auxiliary function introduced for simplifying the integration domain (at the price of making more complex the integrand), namely
\begin{equation}\label{calB}
{S}(q)=
\left\{
\begin{array}{rr}
1, &\qquad q< 0,\\
0, &\qquad q\geq 0.
\end{array}
\right.
\end{equation}

Now, let us introduce also the isotropic symmetry of the collision kernel. Because of this symmetry, the probability density function is further simplified $f(t, \xi)$, for the time $t\in\mathbb{R}^+$ and for the magnitude of the molecular velocity $\xi=\left\|\bm{\xi}\right\|\in\mathbb{R}_\xi^+$. In this way, the distribution function allows one to compute the infinitesimal probability to find some molecules in the time interval between $t$ and $t+dt$ with a velocity magnitude between $\xi$ and $\xi+d\xi$, namely $f(t,\xi)dt d\xi$. Clearly this probability density function can be reformulated in terms of the particle kinetic energy $E=\xi^2/2$, namely $f(t, E)$. Let us introduce the unit vector $\bm{n}_*$ along the direction $\bm{\xi}_*$ and the unit vector $\bm{n}_\odot$ along the direction $\bm{\xi}$, namely
\begin{equation}\label{nstar}
\bm{n}_*=\frac{\bm{\xi}_*}{\left\|\bm{\xi}_*\right\|},\qquad \bm{n}_\odot=\frac{\bm{\xi}}{\left\|\bm{\xi}\right\|}.
\end{equation}
By means of the previous versor, the volume element $d\bm{\xi_*}$ can be expressed as $\xi_*^2 d\bm{n_*}d{\xi_*}$ and consequently
\begin{equation}\label{coll3}
Q(f,f)=\int_{0}^{+\infty}\int_0^{4\pi}\int_0^{4\pi}
\left(f' f_*'-f f_*\right)\,{S}(q)\,B(q)\,\xi_*^2\,
d\bm{n}\,d\bm{n_*}\,d{\xi_*}.
\end{equation}
It is clear that $q$ is the only parameter potentially dependent on directions $\bm{n}$ and $\bm{n}_*$ and, in general,
\begin{equation}\label{q}
q=\bm{\xi}_*\cdot\bm{n}-\bm{\xi}\cdot\bm{n}={\xi}_*\cos{(\alpha_y)}-{\xi}\cos{(\alpha_x)},
\end{equation}
where $\alpha_x$ is the angle between $\bm{\xi}$ and $\bm{n}$, while $\alpha_y$ is the angle between $\bm{\xi}_*$ and $\bm{n}$. Let us introduce the auxiliary variable $x=\cos{(\alpha_x)}$ and $y=\cos{(\alpha_y)}$, namely $q={\xi}_*\,y-{\xi}\,x$. Let us express the surface elements defined by $d\bm{n}$ and $d\bm{n_*}$ in Eq. (\ref{coll3}) by using $\bm{\xi}$ as polar axis for $d\bm{n}$ and $\bm{n}$ as polar axis for $d\bm{n_*}$, namely $d\bm{n}=\sin{(\alpha_x)}\,d\alpha_x\,d\beta_x$ and $d\bm{n_*}=\sin{(\alpha_y)}\,d\alpha_y\,d\beta_y$ respectively, where $\beta_x$ and $\beta_y$ are the corresponding azimuthal angles. This yields
%
%
%
\begin{equation}\label{coll5}
Q(f,f)=\int_{0}^{+\infty}\int_0^{2\pi}\int_{-1}^{+1}\int_0^{2\pi}\int_{-1}^{+1}
\left(f' f_*'-f f_*\right)\,{S}(q)\,B(q)\,\xi_*^2\,
dx\,d\beta_x\,
dy\,d\beta_y\,
d{\xi_*}.
\end{equation}
%
%
Taking the square of Eqs. (\ref{post1}, \ref{post2}) and recalling that $q=\bm{g}\cdot\bm{n}$ yields
\begin{eqnarray}
\label{post1s}
({\xi}')^2&=&{\xi}^2+q^2+2\,q\,{\xi}\,x=
\xi^2\,(1-x^2)+{\xi}_*^2\,y^2,\\
\label{post2s}
({\xi_*}')^2&=&{\xi_*}^2+q^2-2\,q\,\xi_*\,y=
\xi_*^2\,(1-y^2)+{\xi}^2\,x^2.
\end{eqnarray}
Finally, since $\left(f' f_*'-f f_*\right)$ does not depend on $\beta_x$ and $\beta_y$, Eq. (\ref{coll5}) becomes
\begin{equation}\label{coll6}
Q(f,f)=N(f,f)-\nu(f)\,f,
\end{equation}
where
\begin{equation}\label{Nold}
N(f,f)=4\,\pi^2\,\int_{0}^{+\infty}\xi_*^2\,\int_{-1}^{+1}\int_{-1}^{+1}
f(\xi') f(\xi_*')\,{S}(q)\,B(q)\,
dx\,dy\,d{\xi_*},
\end{equation}
\begin{equation}\label{nuold}
\nu(f)=4\,\pi^2\,\int_{0}^{+\infty}f(\xi_*)\,\xi_*^2\,\int_{-1}^{+1}\int_{-1}^{+1}
{S}(q)\,B(q)\,
dx\,dy\,d{\xi_*}.
\end{equation}
The variables $x$ and $y$ are called collisional parameters (integration dummy variables). Let us define $\Omega$ the domain of integration of the collisional parameters, namely $\Omega\dot{=}[-1,1]\times[-1,1]$. It is possible to divide $\Omega$ in two subregions, namely $\Omega_{q\geq0}$ and $\Omega_{q<0}$, defined by $q\geq0$ and $q<0$ respectively. Clearly $\Omega_{q\geq0}\cup\Omega_{q<0}=\Omega$ and they are separated by the condition $q=0$, which is the line $y=\xi/\xi_*\,x$. For any generic point $P\dot{=}(x_P,y_P)\in\Omega_{q<0}$, it is possible to define another point $P_*$ symmetric with regards to the origin, namely $P_*=(-x_P,-y_P)\in\Omega_{q\geq0}$. Taking into account that $B(q)=B(x,y)$, it easy to prove that $B(P)=B(P_*)$ and consequently
\begin{equation}\label{symmetry1}
\int_{-1}^{+1}\int_{-1}^{+1}{S}(q)\,B(q)\,dx\,dy=
\frac{1}{2}\,\int_{-1}^{+1}\int_{-1}^{+1}B(q)\,dx\,dy.
\end{equation}
Recalling Eqs. (\ref{post1s},\ref{post2s}), namely
\begin{eqnarray}
\label{HIpost1s2}
\xi'=\xi'(\xi,\xi_*,x,y)&=&\sqrt{\xi^2\,(1-x^2)+{\xi}_*^2\,y^2},\\
\label{HIpost2s2}
\xi_*'=\xi_*'(\xi,\xi_*,x,y)&=&\sqrt{\xi_*^2\,(1-y^2)+{\xi}^2\,x^2},
\end{eqnarray}
it is easy to prove that $\xi'(P)=\xi'(P_*)$ and $\xi_*'(P)=\xi_*'(P_*)$ and consequently
\begin{equation}\label{symmetry2}
\int_{-1}^{+1}\int_{-1}^{+1}
f(\xi') f(\xi_*')\,{S}(q)\,B(q)\,
dx\,dy=
\frac{1}{2}\,\int_{-1}^{+1}\int_{-1}^{+1}
f(\xi') f(\xi_*')\,B(q)\,
dx\,dy.
\end{equation}
Substituting Eq. (\ref{symmetry2}) into Eq. (\ref{Nold}) and Eq. (\ref{symmetry1}) into Eq. (\ref{nuold}) yields
\begin{equation}\label{N}
N(f,f)=2\,\pi^2\,a^2c_s^{1-\theta}\,\int_{0}^{+\infty}\xi_*^2\,\int_{-1}^{+1}\int_{-1}^{+1}
f(\xi') f(\xi_*')\,\big|{\xi}_*\,y-{\xi}\,x\big|^\theta\,
dx\,dy\,d{\xi_*},
\end{equation}
\begin{equation}\label{nu}
\nu(f)=2\,\pi^2\,a^2c_s^{1-\theta}\,\int_{0}^{+\infty}f(\xi_*)\,\xi_*^2\,\int_{-1}^{+1}\int_{-1}^{+1}
\big|{\xi}_*\,y-{\xi}\,x\big|^\theta\,
dx\,dy\,d{\xi_*}.
\end{equation}

\subsection{Energy formulation}

Let us introduce a change of variables in the previous expressions. Let us introduce $E=\xi^2/2$, $E_*=\xi_*^2/2$, $E'=(\xi')^2/2$ and $E_*'=(\xi_*')^2/2$, namely
\begin{equation}\label{ener_N}
N(f,f) = F c_s^{-\theta} \int_{0}^{+\infty}E_*^{1/2}\,\int_{-1}^{+1}\int_{-1}^{+1}
f(E') f(E_*')\,|y\,E_*^{1/2}-x\,E^{1/2}|^\theta\,
dx\,dy\,d{E_*},
\end{equation}
\begin{equation}\label{ener_nu}
\nu(f) = F c_s^{-\theta} \int_{0}^{+\infty}f(E_*)\,E_*^{1/2}\,\int_{-1}^{+1}\int_{-1}^{+1}
|y\,E_*^{1/2}-x\,E^{1/2}|^\theta\,
dx\,dy\,d{E_*},
\end{equation}
%
%
%
%
where $F=2^{(\theta+3)/2} \pi^2 a^2 c_s$ has the dimensions of a volumetric flow rate. Consequently the collision relations become
\begin{eqnarray}
\label{ener_post1s}
E'&=&E\,(1-x^2)+E_*\,y^2,\\
\label{ener_post2s}
E_*'&=&E\,x^2+E_*\,(1-y^2).
\end{eqnarray}

Let us verify the existence of collisional invariants \cite{cercignani1987} for the previous formulation. Let us introduce the generic macroscopic quantity $\Phi$, namely
\begin{equation}\label{moment}
\Phi(t)=
4\pi\sqrt{2}\int_{0}^{+\infty}\phi(E)\,f\,E^{1/2}\,dE,
\end{equation}
where $\phi(E)$ is a generic function of the particle kinetic energy. The macroscopic dynamics of the quantity $\Phi$ can be computed as
\begin{equation}\label{ener_conserv_Qa}
\frac{d\Phi}{dt}=\int_{0}^{+\infty}Q(f,f)\,\phi(\xi)\,d\bm{\xi} =
4\pi\sqrt{2}\int_{0}^{+\infty}Q(f,f)\,\phi(E)\,E^{1/2}\,dE,
\end{equation}
or equivalently
\begin{equation}\label{moment_eq1}
\frac{d\Phi}{dt}=\left\langle \phi(E),f(E') f(E_*')-f(E) f(E_*) \right\rangle,
\end{equation}
where
\begin{eqnarray}\label{def_inner}
\frac{\left\langle \phi,\varphi \right\rangle}{4\pi\sqrt{2}\,F c_s^{-\theta}}=
 \int_{0}^{+\infty}\int_{0}^{+\infty}\int_{-1}^{+1}\int_{-1}^{+1}
|y\,E_*^{1/2}-x\,E^{1/2}|^\theta\,
\phi\,\varphi\,(E\,E_*)^{1/2}\,dx\,dy\,d{E_*}dE.\nonumber
\end{eqnarray}
Clearly the macroscopic dynamics can not depend on the arbitrary labeling of the microscopic particles. Hence let us invert $E$ and $E_*$ and, since $x=\cos{(\alpha_x)}=\bm{n}_\odot\cdot\bm{n}$ and $y=\cos{(\alpha_y)}=\bm{n}_*\cdot\bm{n}$ (see Eqs. (\ref{nstar})), let us invert the variables $x$ and $y$ as well. Because of these inversions, the following expression holds
\begin{equation}\label{moment_eq2}
\frac{d\Phi}{dt}=\left\langle \phi(E_*),f(E') f(E_*')-f(E) f(E_*) \right\rangle.
\end{equation}
Next, let us invert the pre- and post-collisional velocities.
%
%
%
The collisional parameters expressed by means of the post-collisional velocities become
\begin{eqnarray}\label{xpypa}
x'=\frac{\bm{\xi}'\cdot\bm{n}}{\left\|\bm{\xi}'\right\|}=y\,\sqrt{\frac{E_*}{E\,(1-x^2)+E_*\,y^2}},\\
\label{xpypb}
y'=\frac{\bm{\xi}_*'\cdot\bm{n}}{\left\|\bm{\xi}_*'\right\|}=x\,\sqrt{\frac{E}{E\,x^2+E_*\,(1-y^2)}}.
\end{eqnarray}
It follows immediately that $y'\,\sqrt{E_*'}-x'\,\sqrt{E'}=x\,\sqrt{E}-y\,\sqrt{E_*}$, which ensures that the collisional kernel is unchanged. Equations (\ref{ener_post1s}, \ref{ener_post2s}, \ref{xpypa}, \ref{xpypb}) define the transformation $(E,E_*,x,y)\rightarrow(E',E_*',x',y')$ and they allow one to compute the corresponding Jacobian. Its modulus gives the factor by which the transformation expands or shrinks the infinitesimal volume in the product $\left\langle \phi,\varphi \right\rangle$, namely
\begin{equation}\label{changeofvaria}
dx'\,dy'\,d{E_*'}dE'=\sqrt{\frac{E\,E_*}{E'\,E_*'}}\,dxdy\,dE_*dE.
\end{equation}
Consequently
\begin{equation}\label{moment_eq3}
\frac{d\Phi}{dt}=-\left\langle \phi(E'),f(E') f(E_*')-f(E) f(E_*) \right\rangle.
\end{equation}
%
%
%
By using Eqs. (\ref{moment_eq1}, \ref{moment_eq2}, \ref{moment_eq3}), it is easy to prove \cite{cercignani1987} also for the energy formulation that
\begin{equation}\label{moment_eq4}
\frac{d\Phi}{dt}=\left\langle \phi(E)+\phi(E_*)-\phi(E')-\phi(E_*'),f(E') f(E_*')-f(E) f(E_*) \right\rangle.
\end{equation}
This means that if the quantity $\phi(E)$ is unchanged by the microscopic collision (collisional invariant), then the corresponding macroscopic quantity $\Phi$ is constant in time (conserved quantity). In particular, let us consider the following moments
\begin{equation}\label{momentp}
\Phi_p(t)=
4\pi\sqrt{2}\int_{0}^{+\infty}f\,E^{p+1/2}\,dE,
\end{equation}
which are obtained by taking $\phi(E)=\phi_p=E^p$ into Eq. (\ref{moment}). Clearly $\phi_0=1$ and $\phi_1=E$ are both invariant during the generic microscopic collision: hence, the corresponding macroscopic quantities $\Phi_0$ and $\Phi_1$ are conserved, namely $d\Phi_0/dt=0$ and $d\Phi_1/dt=0$. These macroscopic quantities are usually formulated in terms of number density 
\begin{equation}\label{number}
n=\Phi_0=
4\pi\sqrt{2}\int_{0}^{+\infty}f\,E^{1/2}\,dE,
\end{equation}
and specific internal energy
\begin{equation}\label{energy}
e=\frac{\Phi_1}{\Phi_0}=
\frac{4\pi\sqrt{2}}{n}\int_{0}^{+\infty}f\,E^{3/2}\,dE=
\frac{\int_{0}^{+\infty}f\,E^{3/2}\,dE}{\int_{0}^{+\infty}f\,E^{1/2}\,dE}.
\end{equation}
The collisional invariants $\phi_0=1$ and $\phi_1=E$ (and consequently the conserved quantities $n$ and $e$) are also involved in the definition of the local equilibrium, i.e. the distribution function $f_E$ such that $Q(f_E,f_E)=0$. Let us assume $f_E=\exp[-(c_0\phi_0+c_1\phi_1)]$, where $c_0$ and $c_1$ are some proper constants. The collisional operator $Q(f,f)\propto f' f_{*}'-f f_{*}$ is consequently null, namely
\begin{equation}\label{zerocond}
Q(f_E,f_E)\propto \exp\left[-c_1(E'+E_*')\right]
-\exp\left[-c_1(E+E_*)\right]=0.
\end{equation}
The constants $c_0$ and $c_1$ can be found by ensuring that Eqs. (\ref{number}, \ref{energy}) are satisfied, namely
\begin{equation}\label{eq_E}
f_E=\frac{n}{(2\pi E_B)^{3/2}}\,\exp{\left(-\frac{E}{E_B}\right)},
\end{equation}
where $E_B=2 e/3$. Recalling that the pressure $P$ is defined as one third of the stress tensor trace \cite{cercignani1987}, it follows that $P=2/3\,n\,e=n\,E_B$. Moreover, recalling the ideal gas law, i.e. $P=n\,k_B\,T$, where $k_B$ is the Boltzmann constant and $T$ is the temperature, it follows that $2e/3=E_B=k_B\,T$. Introducing the specific heat capacity (per mole) at constant volume $C_v=e/T$, it follows that $C_v=3/2\,k_B$, which is correct for monatomic gases considered here.

\subsection{Hierarchy of moment equations}

Sometimes it is more convenient to compute Eq. (\ref{moment_eq1}) in a slightly different way, namely
\begin{equation}\label{moment_eq1b}
\frac{d\Phi}{dt}=\left\langle \phi(E),f(E') f(E_*')\right\rangle
-\left\langle \phi(E),f(E) f(E_*)\right\rangle.
\end{equation}
It has already been shown (in the previous section) that the product $\left\langle \phi,\varphi \right\rangle$ can be equivalently formulated in terms of the post-collisional velocities $\left\langle \phi,\varphi \right\rangle'$ (since the collisional kernel is invariant and the infinitesimal volume can be transformed by Eq. (\ref{changeofvaria})). In particular, once the inverse transformation $(E',E_*',x',y')\rightarrow(E,E_*,x,y)$, namely
\begin{eqnarray}
\label{inverse_e}
E&=&E'\,[1-(x')^2]+E_*'\,(y')^2,\\
\label{inverse_es}
E_*&=&E'\,(x')^2+E_*'\,[1-(y')^2],\\
\label{inverse_x}
x&=&y'\,\sqrt{\frac{E_*'}{E'\,[1-(x')^2]+E_*'\,(y')^2}},\\
\label{inverse_y}
y&=&x'\,\sqrt{\frac{E'}{E'\,(x')^2+E_*'\,[1-(y')^2]}},
\end{eqnarray}
is used for evaluating $\phi(E)=\phi(E(E',E_*',x',y'))$, the first term in the right hand side of Eq. (\ref{moment_eq1b}) can be rewritten as
\begin{equation}\label{term}
\left\langle \phi(E),f(E') f(E_*')\right\rangle=
\left\langle \phi(E(E',E_*',x',y')),f(E') f(E_*')\right\rangle'.
\end{equation}
Omitting the prime symbol in the previous expression allows one to reformulate Eq. (\ref{moment_eq1b}) as
\begin{equation}\label{moment_eq1c}
\frac{d\Phi}{dt}=\left\langle \phi(E\,(1-x^2)+E_*\,y^2)
-\phi(E),f(E) f(E_*)\right\rangle.
\end{equation}
The previous equation is usually the starting point of the so-called quadrature method of moments (QMOM), which is a generic solution method for population balance models \cite{marchisio2005}.

%

\subsection{Numerical integration of energy formulation}

Let us assume a maximum value for the test particle kinetic energy $E$, namely $E_M$. Let us divide the interval $[0,\,E_M]$ in $M$ equal parts, with length $\Delta E=E_M/M$. Each cell is identified by index $1\leq i\leq M$, such that $E_i=(i-1/2)\,\Delta E$, and the probability distribution function is discretized accordingly, namely $f_i=f(E_i)$. As suggested by Ref. \cite{aristov2001}, this simple discretization (piecewise constant) can be used to compute a numerical approximation $\tilde{\nu}_i$ of the relaxation frequency $\nu_i$ for the discrete probability distribution function $f_i$, namely
\begin{equation}\label{e_nu_num}
\nu_i=\nu(f_i)\approx\tilde{\nu}_i = \tilde{F}\,\Delta{E}\,\sum_{j=1}^M f_j\,E_j^{1/2}\,A_{ij},
\end{equation}
where 
%
\begin{equation}\label{e_A}
A_{ij} = \Delta E^{-\theta/2} \int_{-1}^{+1}\int_{-1}^{+1}
\left|y\,E_j^{1/2}-x\,E_i^{1/2}\right|^\theta\,
dx\,dy.
\end{equation}
and 
\begin{equation}\label{tildeF}
\tilde{F}=F\,\left(\frac{\sqrt{\Delta E}}{c_s}\right)^\theta=2^{(\theta+3)/2} \pi^2 a^2 c_s\,\left(\frac{\sqrt{\Delta E}}{c_s}\right)^\theta.
\end{equation}
%
%
The previous expression admits analytical solution, namely
\begin{equation}\label{e_genA}
A_{ij}(\theta) = 2\,\Delta E^{-\theta/2} 
\frac{\left|E_i^{1/2}+E_j^{1/2}\right|^{2+\theta}-\left|E_i^{1/2}-E_j^{1/2}\right|^{2+\theta}}
{E_i^{1/2}E_j^{1/2}(2+3\theta+\theta^2)},
\end{equation}
which for $\theta=0$ (constant kernel model) yields $A_{ij}(0)=4$, while for $\theta=1$ (hard sphere model, consistent with Ref. \cite{aristov2001}) yields
\begin{equation}\label{e_genAaristov}
A_{ij}(1) = \frac{1}{\sqrt{\Delta E}} \left\{
\begin{array}{rr}
2\,E_j^{1/2}+2/3\,E_i\,E_j^{-1/2}, &\qquad E_i\leq E_j,\\
2\,E_i^{1/2}+2/3\,E_j\,E_i^{-1/2}, &\qquad E_i>E_j.
\end{array}
\right.
\end{equation}
%
%
Similarly, the piecewise discretization can be used to compute a numerical approximation $\tilde{N}_i$ of the relaxation frequency $N_i$ for the discrete probability distribution function $f_i$, namely
\begin{equation}\label{e_N_num}
N_i=N(f_i,f_i)\approx \tilde{N}_i=\tilde{F}\,\Delta{E}\,\sum_{j=1}^M E_j^{1/2}\,\sum_{k=1}^M\sum_{l=1}^M f_k f_l\,B_{ij}^{kl},
\end{equation}
where
%
%
%
%
\begin{equation}\label{e_B}
B_{ij}^{kl}=
\Delta E^{-\theta/2}\int_{\Omega_{ij}^{kl}}
\left|y\,E_j^{1/2}-x\,E_i^{1/2}\right|^\theta\,
dx\,dy,
\end{equation}
and $\Omega_{ij}^{kl}$ is the compatibility domain (which may also be null). The domain $\Omega_{ij}^{kl}$ is defined as the locus of points $(x,y)\in\Omega$ such that the post-collisional energies $\tilde{E}'(x,y)$ and $\tilde{E}_*'(x,y)$, defined as
\begin{eqnarray}
\label{ener_post1sb}
\tilde{E}'(x,y)&=&E_i\,(1-x^2)+E_j\,y^2,\\
\label{ener_post2sb}
\tilde{E}_*'(x,y)&=&E_j\,(1-y^2)+E_i\,x^2,
\end{eqnarray}
are approximated by (piecewise) constants over a small region around the point $(E_k,E_l)$. Let us define by $E_{k-}=(k-1)\Delta E$ the lower rounding limit and by $E_{k+}=k\Delta E$ the higher rounding limit (similarly for $E_{l-}$ and $E_{l+}$). Consequently the pair $(x,y)$ belongs to $\Omega_{ij}^{kl}$ if $(\tilde{E}',\tilde{E}_*')$ belongs to $[E_{k-},\,E_{k+}]\times[E_{l-},\,E_{l+}]$, or equivalently
\begin{eqnarray}
\label{e_post1sb}
E_{k-}&\leq
E_i\,(1-x^2)+E_j\,y^2&\leq E_{k+},\\
\label{e_post2sb}
E_{l-}&\leq
E_j\,(1-y^2)+E_i\,x^2&\leq E_{l+}.
\end{eqnarray}
Taking into account that $\tilde{E}_*'=\tilde{E}_*'(\tilde{E}')=E_i+E_j-\tilde{E}'$, only a segment of the function $\tilde{E}_*'=\tilde{E}_*'(\tilde{E}')$ can (diagonally) fit into the surface element. Hence, in order to define $\Omega_{ij}^{kl}$, it is enough to solve Eq. (\ref{e_post1sb}), which can be reformulated as 
\begin{eqnarray}
\label{e_hyperbola1base}
\Omega_+&=&\{(x,y)\in\Omega : E_i\,(1-x^2)+E_j\,y^2\leq E_{k+}\},\\
\label{e_hyperbola2}
\Omega_{-\infty}&=&\{(x,y)\in\Omega : E_i\,(1-x^2)+E_j\,y^2\geq E_{k-}\},\\
\Omega_{ij}^{kl}&=&\Omega_+\cap\Omega_{-\infty}.
\end{eqnarray}
The regions $\Omega_+$ and $\Omega_{-\infty}$ are bounded by two hyperbolas and the region $\Omega_{ij}^{kl}$ is the generic intersection between them. This way of defining $\Omega_{ij}^{kl}$ is not efficient because it requires two different formulas for defining $\Omega_+$ and $\Omega_{-\infty}$ respectively. However the problem can be conveniently reformulated, namely
\begin{eqnarray}
\label{e_hyperbola2base}
\Omega_{-}&=&\{(x,y)\in\Omega : E_i\,(1-x^2)+E_j\,y^2\leq E_{k-}\},\\
\Omega_{ij}^{kl}&=&\Omega_+-\Omega_{-}.
\end{eqnarray}
In this way, $\Omega_+$ and $\Omega_{-}$ are defined by the same formula and similarly for the integrals defined over them, which can be computed by a unique expression, namely
\begin{equation}\label{e_Bo}
B_{ij}^{kl}=C(E_{k+})-C(E_{k-}),
\end{equation}
where
\begin{equation}\label{Cpm}
C(E_{k\pm})=
\Delta E^{-\theta/2}\int_{\Omega_{\pm}(E_{k\pm})}
\left|y\,E_j^{1/2}-x\,E_i^{1/2}\right|^\theta\,
dx\,dy.
\end{equation}
In particular, the shape of the domains $\Omega_{\pm}$ on the plane $(x,y)$ depend on the relative magnitude of the energies $E_i$, $E_j$, $E_{k-}$ and $E_{k+}$. As it will be discussed in the next subsections, six cases are possible, but only three formulas ($C_1$, $C_2$ and $C_3$) are required by conveniently switching the arguments, namely
\begin{equation}\label{Cpmgeneric}
C(E_{k\pm}) = \left\{
\begin{array}{rr}
C_1(E_i,E_j,E_{k\pm}), &\qquad E_{k\pm}\leq E_i\leq E_j,\\
C_2(E_i,E_j,E_{k\pm}), &\qquad E_i\leq E_{k\pm}\leq E_j,\\
C_3(E_i,E_j,E_{k\pm}), &\qquad E_i\leq E_j\leq E_{k\pm},\\
C_1(E_j,E_i,E_{k\pm}), &\qquad E_{k\pm}\leq E_j<E_i,\\
C_2(E_j,E_i,E_{k\pm}), &\qquad E_j\leq E_{k\pm}\leq E_i,\\
C_3(E_j,E_i,E_{k\pm}), &\qquad E_j<E_i\leq E_{k\pm}.
\end{array}
\right.
\end{equation}
Hence, in the following subsections, only the first three cases are discussed.

\subsubsection{Case 1: $E_{k\pm}\leq E_i\leq E_j$}

In this case, the domain $\Omega_{\pm}$ is defined by
\begin{eqnarray}
\label{e_hyperbola1b}
\frac{x^2}{a_\pm^2}-\frac{y^2}{b_\pm^2}&\geq&1,
\end{eqnarray}
where $a_\pm=\sqrt{1-E_{k\pm}/E_i}$ and $b_\pm=\sqrt{\left(E_i-E_{k\pm}\right)/E_j}$. The domain $\Omega_\pm$ is simply made of two strips between two hyperbolas. Taking into account the already discussed symmetry of the problem with regards to the origin of the plane $(x,y)$, it is possible to save some computations. In particular, considering only the strip such that $q(E_i,E_j)\leq 0$, the function $C_1(E_i,E_j,E_{k\pm})$ can be expressed as
\begin{equation}\label{e_C1}
C_1(E_i,E_j,E_{k\pm})=
2\,\Delta E^{-\theta/2}\int_{a_\pm}^{+1}
\int_{-c_\pm(x)}^{c_\pm(x)}
\,\left|y\,E_j^{1/2}-x\,E_i^{1/2}\right|^\theta\,
dy\,dx,
\end{equation}
where 
\begin{equation}\label{e_C}
c_\pm(x)=\sqrt{\left(E_i\,x^2-E_i+E_{k\pm}\right)/E_j}.
\end{equation}
The previous expression $C_1=C_1(\theta)$ can be found analytically for particular values of $\theta\in\mathbb{N}$ (a generic expression in terms of $\theta$ was not found). In particular, for $\theta=0$ (constant kernel model) 
\begin{eqnarray}\label{e_C1bt0}
C_1(0)=
2\,\int_{a_\pm}^{+1}
\int_{-c_\pm(x)}^{c_\pm(x)}
\left(E_i^{1/2}\,x-E_j^{1/2}\,y\right)\,dy\,dx=\nonumber\\	
2\,\frac{E_{k\pm}^{1/2}}{E_j^{1/2}}+\frac{E_i-E_{k\pm}}{E_i^{1/2} E_j^{1/2}}\,
\ln\left(\frac{E_i^{1/2}-E_{k\pm}^{1/2}}{E_i^{1/2}+E_{k\pm}^{1/2}}\right),
\end{eqnarray}
and for $\theta=1$ (hard spheres model)
\begin{equation}\label{e_C1bt1}
C_1(1)=\frac{2}{{\Delta E}^{1/2}}\,
\int_{a_\pm}^{+1}
\int_{-c_\pm(x)}^{c_\pm(x)}
\left(E_i^{1/2}\,x-E_j^{1/2}\,y\right)\,
dy\,dx=\frac{4}{3\,{\Delta E}^{1/2}}\,\frac{E_{k\pm}^{3/2}}{E_i^{1/2}\,E_j^{1/2}}.
\end{equation}
%
%
%
From the previous expressions, if $E_{k-}$ ($<E_{k+}$) is minimum, i.e. $E_{k-}=0$, then $C_1(0)=C_1(1)=0$. 

\subsubsection{Case 2: $E_i\leq E_{k\pm}\leq E_j$}

In this case, the domain $\Omega_\pm$ is defined by
\begin{eqnarray}
\label{e_hyperbola1c}
\frac{y^2}{b_\pm^2}-\frac{x^2}{a_\pm^2}&\leq&1,
\end{eqnarray}
where $a_\pm=\sqrt{E_{k\pm}/E_i-1}$ and $b_\pm=\sqrt{\left(E_{k\pm}-E_i\right)/E_j}$. The domain $\Omega_\pm$ is again made of two strips between two hyperbolas, but the function $C_2(E_i,E_j,E_{k\pm})$ can be computed by means of one strip only ($q(E_i,E_j)\leq 0$). The integral $C_2$ over $\Omega_\pm$ depends on the coordinates $(x_I,y_I)$ of the intersections between the previous hyperbolas and $y_I=\pm 1$. The abscissas of these intersections are $x_I=\pm e_\pm$ where
\begin{equation}\label{e_E}
e_\pm=
\sqrt{1+\frac{E_j-E_{k\pm}}{E_i}}.
\end{equation}
In particular, for $E_j\geq E_{k\pm}$, which is the present case, $e_\pm\geq 1$ and consequently the intersections $(x_I,y_I)$ are out of the domain $\Omega_\pm$. Hence, for the preset case, we can neglect this problem. Consequently 
\begin{equation}\label{e_C2}
C_2(E_i,E_j,E_{k\pm})=
2\,\Delta E^{-\theta/2}\,\int_{-1}^{+1}
\int_{-c_\pm(x)}^{d(x)}
\left(E_i^{1/2}\,x-E_j^{1/2}\,y\right)^\theta\,
dy\,dx,
\end{equation}
where $c_\pm(x)$ is defined by Eq. (\ref{e_C}) and $d(x)=x\,\sqrt{E_i/E_j}$. The previous integral $C_2=C_2(\theta)$ admits analytical solutions for particular values of $\theta\in\mathbb{N}$. In particular, for $\theta=0$ (constant kernel model) 
\begin{equation}\label{e_C2bt0}
C_2(0)=2\,\frac{E_{k\pm}^{1/2}}{E_j^{1/2}}+\frac{E_{k\pm}-E_i}{E_i^{1/2} E_j^{1/2}}\,
\ln\left(\frac{E_{k\pm}^{1/2}+E_i^{1/2}}{E_{k\pm}^{1/2}-E_i^{1/2}}\right),
\end{equation}
and for $\theta=1$ (hard spheres model)
\begin{equation}\label{e_C2bt1}
C_2(1)=2\,\frac{3\,E_{k\pm}-E_i}{3\,{\Delta E}^{1/2}\,E_j^{1/2}}.
\end{equation}
%

\subsubsection{Case 3: $E_i\leq E_j\leq E_{k\pm}$}

In this case, the domain $\Omega_\pm$ is defined by
\begin{eqnarray}
\label{e_hyperbola1d}
\frac{y^2}{b_\pm^2}-\frac{x^2}{a_\pm^2}&\leq&1,
\end{eqnarray}
where $a_\pm=\sqrt{E_{k\pm}/E_i-1}$ and $b_\pm=\sqrt{\left(E_{k\pm}-E_i\right)/E_j}$. The domain $\Omega_\pm$ is made of a combination of two hyperbolas and the boundaries of $\Omega$: however it is still symmetric with regards to the origin and hence the function $C_3(E_i,E_j,E_{k\pm})$ can be expressed by means of the subregion with $q(E_i,E_j)\leq 0$. Since $E_j\leq E_{k\pm}$, $e_\pm\leq 1$ where $e_\pm$ is given by Eq. (\ref{e_E}) and consequently the intersections $(x_I,y_I)$ between the previous hyperbolas and $y_I=\pm 1$ are inside the domain $\Omega_\pm$. Consequently 
\begin{eqnarray}\label{e_C3}
C_3(E_i,E_j,E_{k\pm})=
&&2\,\Delta E^{-\theta/2}\,\int_{-1}^{-e_\pm}
\int_{-1}^{d(x)}
\left(E_i^{1/2}\,x-E_j^{1/2}\,y\right)^\theta\,
dy\,dx+\nonumber\\
&&2\,\Delta E^{-\theta/2}\,\int_{-e_\pm}^{+e_\pm}
\int_{-c_\pm(x)}^{d(x)}
\left(E_i^{1/2}\,x-E_j^{1/2}\,y\right)^\theta\,
dy\,dx+\nonumber\\
&&2\,\Delta E^{-\theta/2}\,\int_{+e_\pm}^{+1}
\int_{-1}^{d(x)}
\left(E_i^{1/2}\,x-E_j^{1/2}\,y\right)^\theta\,
dy\,dx,
\end{eqnarray}
where $c_\pm(x)$ is given by Eq. (\ref{e_C}), $d(x)=x\,\sqrt{E_i/E_j}$ and $e_\pm$ is given by Eq. (\ref{e_E}). The previous integral $C_3=C_3(\theta)$ admits an analytical solution for particular values of $\theta\in\mathbb{N}$. In particular, for $\theta=0$ (constant kernel model) 
\begin{eqnarray}\label{e_C3bt0}
C_3(0)&=&4-2\,\sqrt{\frac{E_i+E_j-E_{k\pm}}{E_i}}\nonumber\\
&&-2\,\frac{E_i-E_{k\pm}}
{E_i^{1/2} E_j^{1/2}}\,
\ln\left[\frac{E_j^{1/2}+\left(E_i+E_j-E_{k\pm}\right)^{1/2}}
{\left(E_j-(E_i+E_j-E_{k\pm})\right)^{1/2}}\right],
\end{eqnarray}
and for $\theta=1$ (hard spheres model)
\begin{equation}\label{e_C3bt1}
C_3(1)=\frac{2}{{\Delta E}^{1/2}}\left[E_j^{1/2}+\frac{1}{3}\,\frac{E_i}{E_j^{1/2}}
-\frac{2}{3\,E_i^{1/2}E_j^{1/2}}\,\left(E_i+E_j-E_{k\pm}\right)^{3/2}\right].
\end{equation}
Clearly, the previous expressions are always well defined, because the maximum value of $E_{k+}$ ($>E_{k-}$) is exactly $E_{k+}=E_i+E_j$, which corresponds to $E_{l-}=0$. In particular, if $E_{k+}=E_i+E_j$, then $C_3(0)=A_{ij}(0)$ and $C_3(1)=A_{ij}(1)$. 


\subsection{\label{DVM}Discrete Velocity Model (DVM) and master equation}

As already pointed out, the compatibility domain $\Omega_{ij}^{kl}$ is defined as the locus of points $(x,y)\in\Omega$ such that, for some given pre-collisional energies $(E_i,E_j)$, the post-collisional energies $\tilde{E}'(x,y)$ and $\tilde{E}_*'(x,y)$ (see Eqs.(\ref{ener_post1sb}, \ref{ener_post2sb})) are in the neighborhood of the node $(E_k,E_l)$ (coherently with the adopted piecewise approximation). Clearly the compatibility domain may also be null. In particular, two cases may be distinguished. If the pre-collisional energies are such that $E_i+E_j\leq E_M$, then all the post-collisional energies fit into the adopted discretization mesh for the kinetic energy. On the other hand, if $E_i+E_j>E_M$, then some post-collisional energies are still physically possible, but they fall outside the discretization mesh (and they should be excluded for consistency). Hence purely geometrical considerations yield the following property, namely
\begin{eqnarray}\label{domain-decom}
\bigcup_{k=1}^M\bigcup_{l=1}^M \Omega_{ij}^{kl}&=&\Omega,\qquad\mbox{for}\qquad E_i+E_j\leq E_M,\\
\bigcup_{k=1}^M\bigcup_{l=1}^M \Omega_{ij}^{kl}&<&\Omega,\qquad\mbox{for}\qquad E_i+E_j>E_M,
\end{eqnarray}
%
%
%
and consequently
\begin{eqnarray}\label{e_AvB}
\sum_{k=1}^M\sum_{l=1}^M B_{ij}^{kl}&=&A_{ij},\qquad\mbox{for}\qquad E_i+E_j\leq E_M,\\
\sum_{k=1}^M\sum_{l=1}^M B_{ij}^{kl}&<&A_{ij},\qquad\mbox{for}\qquad E_i+E_j>E_M.
\end{eqnarray}
In case $E_i+E_j\leq E_M$, the fact that the equality is exactly satisfied (by the discrete numerical operators) is a consequence of the energy formulation, which allows one to ensure perfect conservation of particle number and energy on a discrete lattice. On the other hand, if $E_i+E_j>E_M$, the pre-collisional energies starting from outside of the discretization mesh are automatically excluded (even though they are physically possible) and hence also the post-collisional energies falling outside the discretization mesh should be excluded as well for consistency. In this way, all the direct and reverse collisions live on the same discretization mesh. The latter strategy is advantageous from the computational point of view, but it reveals that the adopted numerical description only approximates the dynamics due to the collisions with $E_i+E_j>E_M$. In case that very accurate simulations are required, it would be better to focus on the sub-region $[0,E_M/2]$.

Let us define the following matrix
\begin{equation}\label{e_AvB2}
\hat{A}_{ij}= \sum_{k=1}^M\sum_{l=1}^M B_{ij}^{kl},
\end{equation}
and consequently Eq. (\ref{e_nu_num}) becomes
\begin{equation}\label{e_nu_num2}
\nu_i \approx\tilde{\nu}_i= F\,\Delta{E}\,\sum_{j=1}^M f_j\,E_j^{1/2}\,\hat{A}_{ij}.
\end{equation}
Introducing $\tilde{Q}_i=\tilde{N}_i-\tilde{\nu}_i f_i$ and taking into account Eqs. (\ref{e_N_num}, \ref{e_nu_num2}) yield
\begin{equation}\label{e_coll7}
Q_i=Q(f_i,f_i)\approx\tilde{Q}_i= 
\tilde{F}\,\Delta{E}\,\sum_{j=1}^M E_j^{1/2}\,\left(\sum_{k=1}^M\sum_{l=1}^M f_k f_l\,B_{ij}^{kl}
-f_i f_j\,\hat{A}_{ij}\right).
\end{equation}
We would like to investigate the (macroscopic) conservation properties of the discrete operator $\tilde{Q}_i$. In order to do this, let us rewrite Eq. (\ref{changeofvaria}) for the discrete case, which is simplified by the fact that $dE_*'$, $dE'$, $dE_*$ and $dE$ are all approximated by $\Delta E$, namely
\begin{equation}\label{changeofvaria_d}
dx'\,dy'=\sqrt{\frac{E_i\,E_j}{\tilde{E}'(x,y)\,\tilde{E}_*'(x,y)}}\,dx\,dy\approx\sqrt{\frac{E_i\,E_j}{E_k\,E_l}}\,dx\,dy.
\end{equation}
Clearly the last relation is only asymptotically satisfied by the discrete operator: hence, even though the microscopic collisions are conservative (in terms of mass and kinetic energy), the corresponding macroscopic moments are not exactly conserved. The previous relation suggest to multiply and divide Eq. (\ref{e_coll7}) by $E_i^{1/2}$, which allows one to recover the underlaying Discrete Velocity Model (DVM) \cite{gatignol1975}\footnote{In Eq. (\ref{e_coll9}), we have adopted a dimensionless $\Gamma_{ij}^{kl}$, which is different from the convention used in Ref. \cite{gatignol1975}. However we note that $[F]\,[E^{3/2}]\,[f]=[s]^{-1}$, where $[\cdot]$ means the physical dimensions. Another difference with regards to Ref. \cite{gatignol1975} is due to the term $E_i^{1/2}$ at the denominator, because of the homogeneous isotropic formulation considered here.}, namely
\begin{equation}\label{e_coll9}
\frac{\partial f_i}{\partial t}= \frac{F\Delta E^2}{E_i^{1/2}}\sum_{j,k,l=1}^M \Gamma_{ij}^{kl}\,\left(f_k f_l-f_i f_j\right),
\end{equation}
where
\begin{equation}\label{e_coll8}
\Gamma_{ij}^{kl}= \frac{\sqrt{E_i\,E_j}}{\Delta E}\,B_{ij}^{kl}.
\end{equation}
The following properties hold \cite{gatignol1975}, namely
\begin{equation}\label{prop_sym}
\Gamma_{ij}^{kl}=\Gamma_{ji}^{kl}\approx\Gamma_{kl}^{ij}.
\end{equation}
We would like to mention that this kind of models may be affected by the problem of (spurious) conservation laws \cite{bobilev2008}. In this particular case, numerical meshes in the velocity/energy space (i.e. lattices) large enough should fix the problem from the practical point of view. Equation (\ref{e_coll9}) is sometimes also called master equation and $\Gamma_{ij}^{kl}$ is called the matrix of transition frequencies.

It is possible to correct the matrix of transition frequencies such that it satisfies exactly the symmetry properties (DVM correction). There are $24=4!$ possible permutations of the four indexes $i$, $j$, $k$ and $l$ in the matrix of transition frequencies, but only eight permutations ensure the conservation of kinetic energy. If two indexes are equal ($i=j$ or $k=l$), then only four permutations (conserving kinetic energy) are possible. Let us define by $\{\Gamma_{ij}^{kl}\}$ the set of transition frequencies obtained by permutations of the indexes $(i,j,k,l)$ conserving kinetic energy. The DVM correction is defined as
\begin{equation}\label{correction}
\forall\,(i,j,k,l):\Gamma_{ij}^{kl}\in\{\Gamma_{ij}^{kl}\}, \qquad \tilde{\Gamma}_{ij}^{kl}=\overline{\{\Gamma_{ij}^{kl}\}},
\end{equation}
where the overline means the arithmetic mean of the considered set (symmetrization). By means of this DVM correction, the following property holds (exactly)
\begin{equation}\label{prop_sym_corr}
\tilde{\Gamma}_{ij}^{kl}=\tilde{\Gamma}_{ji}^{kl}=\tilde{\Gamma}_{kl}^{ij},
\end{equation}
as required by the DVM models \cite{gatignol1975}. Consequently it is possible to correct the dimensionless frequencies, namely
\begin{equation}\label{e_coll_corr}
\tilde{B}_{ij}^{kl}= \frac{\Delta E}{\sqrt{E_i\,E_j}}\,\tilde{\Gamma}_{ij}^{kl},
\end{equation}
\begin{equation}\label{e_AvB2b}
\tilde{A}_{ij}= \sum_{k=1}^M\sum_{l=1}^M \tilde{B}_{ij}^{kl},
\end{equation}
which ensure that both particle number and kinetic energy are perfectly conserved also at macroscopic level. In the following, the symbols $\tilde{\nu}_i$ and $\tilde{Q}_i$ are still used (for keeping the notation as simple as possible), even thought they are computed by $\tilde{B}_{ij}^{kl}$ and $\tilde{A}_{ij}$ instead of $B_{ij}^{kl}$ and $\hat{A}_{ij}$. It is worth the effort to point out that, because of the DVM correction, $\tilde{A}_{ij}\neq A_{ij}$ even for $E_i+E_j\leq E_M$ (while, under the same conditions, $\hat{A}_{ij}=A_{ij}$).

Ensuring the numerical conservation of conserved hydrodynamic moments is also one of the key ideas behind the derivation of the so-called Lattice Boltzmann Method (LBM) \cite{succi2001} (even though mass and momentum only are conserved on the smallest lattices).


\subsection{\label{quad}Quadrature formulas for computing the moments}

In order to compute the moments defined by Eq. (\ref{momentp}), the piecewise constant approximation is used. This is consistent with the receipt used for solving the collisional integral $Q(f,f)=N(f,f)-\nu\,f$. It is worth the effort to point out that the property given by Eq. (\ref{prop_sym_corr}) (and ensured numerically by means of the DVM correction) implies the conservation of particle number and energy, only if the piecewise constant approximation is used. Hence, even though more elaborate quadrature formulas are possible for computing the moments, they would spoil the main advantage of the DVM correction, i.e. ensuring that the conservation laws are perfectly satisfied. According to the piecewise constant approximation, Eq. (\ref{momentp}) can be approximated by
\begin{equation}\label{momentp2}
\Phi_p\approx\tilde{\Phi}_p=
4\pi\sqrt{2}\,\Delta E\,\sum_{i=1}^M f_i\,E_i^{p+1/2}.
\end{equation}
This way of computing the moments is straightforward, but it produces some problems in defining the local equilibrium. Let us suppose to define the local discrete equilibrium as $(f_E)_i=f_E(E_i)$, i.e. the local discrete equilibria coincide with the nodal values of the continuous function $f_E$ defined by Eq. (\ref{eq_E}) (for some values of $n$ and $e$). Applying the previous definition yields $\tilde{\Phi}_0((f_E)_i)\neq n$ and $\tilde{\Phi}_1((f_E)_i)\neq n\,e$, where $n$ and $e$ are defined by continuous integrals in Eq. (\ref{number}) and Eq. (\ref{energy}) respectively. This is clearly an effect of the numerical error due to the quadrature formula. 

In order to circumvent this problem, let us define the local equilibrium in the following way by recursive tuning. For any discrete distribution function $f_i$, let us define $\tilde{n}=\tilde{\Phi}_0(f_i)$ and $\tilde{n}\,\tilde{e}=\tilde{\Phi}_1(f_i)$. Let us define the 
\begin{equation}\label{eq_E_num}
(\tilde{f}_E)_i=\exp[-(\tilde{c}_0+\tilde{c}_1\,E_i)],
\end{equation}
where the constants $\tilde{c}_0$ and $\tilde{c}_1$ are defined such that
\begin{equation}\label{eq_E_num_cond}
\tilde{\Phi}_0\left((\tilde{f}_E)_i\right)=\tilde{n},\qquad
\tilde{\Phi}_1\left((\tilde{f}_E)_i\right)=\tilde{n}\,\tilde{e}.
\end{equation}
By means of this recursive tuning of the local discrete equilibrium, the particle number and energy are both constant during the whole relaxation process. Eventually, if the continuous distribution function is known as initial condition, the assumptions $\tilde{n}={\Phi}_0(f)$ and $\tilde{n}\,\tilde{e}={\Phi}_1(f)$ (by Eq. (\ref{momentp})) can be used instead.

\subsection{\label{sBGK}Recovering BGK}

It is well known that the collisional integral of the Boltzmann equation drives any initial distribution function towards the local equilibrium \cite{cercignani1987}. When the distribution function is very close to the local equilibrium, the remaining dynamics becomes very slow (on the kinetic time scale) and it can be described by the so-called fluid dynamic time scale (which is suitable for describing phenomena in the corresponding fluid dynamic regime). Let us search for simplified expressions of the collisional integral $\tilde{Q}_i$ in such regime. The key idea is to use the equilibrium distribution function for computing an approximation of the relaxation frequency given by Eq. (\ref{e_nu_num2}). Since we search for an approximation of the real relaxation frequency, let us consider $A_{ij}$ (admitting analytical expression) instead of $\hat{A}_{ij}$ in Eq. (\ref{e_nu_num2}), namely $\tilde{\nu}_i\approx(\nu_E)_i$ where
\begin{equation}\label{nue}
(\nu_E)_i= 
\tilde{F}\,\Delta{E}\,\sum_{j=1}^M A_{ij}\,E_j^{1/2}\,(\tilde{f}_E)_j.
\end{equation}
Consequently, recalling Eq. (\ref{e_coll7}), it is possible to introduce the following approximation
\begin{equation}\label{BGKlike}
\tilde{Q}_i\approx(\tilde{Q}_{B})_i = (\nu_E)_i\,\left[(\tilde{f}_E)_i-f_i\right].
\end{equation}
In general, $(\nu_E)_i$ still depends on the particle kinetic energy $E_i$. For fixing the ideas, let us consider the Constant Kernel Model - CKM ($\theta=0$ in Eq. (\ref{generic})), where $A_{ij}(0)=4$ and
\begin{equation}\label{nue_ckm}
\nu_{E}(0)= 
4\,\tilde{F}\,\Delta{E}\,\sum_{j=1}^M E_j^{1/2}\,(\tilde{f}_E)_j=
\frac{\tilde{F}\,\tilde{n}}{\pi\sqrt{2}}=\frac{F\,\tilde{n}}{\pi\sqrt{2}},
\end{equation}
i.e. the approximated relaxation frequency $\nu_{E}(0)$ is a constant which depends on the local number density. A similar procedure can be followed for the Hard Sphere Model - HSM ($\theta=1$ in Eq. (\ref{generic})), which also admits an analytical expression for $(\nu_{E})_i(1)$ involving the error function: see Ref. \cite{cercignani1987} for details. For the present purposes, i.e. the discussion of the numerical results of the test case, let us derive the limit of $(\nu_{E})_i(1)$ for high kinetic energies (by considering the case $E_i>E_j$ in Eq. (\ref{e_genAaristov})), namely
\begin{equation}\label{nue_hsm1}
\lim_{E_i\rightarrow E_M}(\nu_{E})_i(1)\approx 
\frac{2\,E_i^{1/2}}{c_s}\,F\,\Delta{E}\,\sum_{j=1}^M E_j^{1/2}\,(\tilde{f}_E)_j=
\frac{F\,\tilde{n}}{\pi\sqrt{2}}\,\left(\frac{\sqrt{E_i}}{2\,c_s}\right).
\end{equation}

From the previous limiting case, it is possible to derive the so-called BGK approximation \cite{cercignani1987}, where BGK stands for Bhatnagar-Gross-Krook who proposed this simple collisional model. The key idea is to assume a constant relaxation frequency (depending on the local number density), namely
\begin{equation}\label{BGK}
(\tilde{Q}_{BGK})_i = \nu_{BGK}\,\left[(\tilde{f}_E)_i-f_i\right].
\end{equation}
In the following, we will assume for simplicity $\nu_{BGK}=(\nu_E)_1$ where $(\nu_E)_1$ stands for $(\nu_E)_i$ at $E_1=\Delta E/2$, even though it should be (more precisely) $\nu_{BGK}=\lim_{E\rightarrow 0}\nu_E(E)$.

\section{\label{overview}Overview of the software structure}

In this section, we provide an overview of the \code{HOMISBOLTZ} program which was developed using Matlab\textregistered. The basic idea is to provide a simple illustration of the discussed methodology, which can be easily ported to other environments (FORTRAN, C++,...). The \code{HOMISBOLTZ} program is free software, which can be redistributed and/or modified under the terms of the GNU General Public License. The \code{HOMISBOLTZ} program has been purposely designed in order to be minimal, not only with regards to the reduced number of lines (less than 1,000), but also with regards to the coding style (as simple as possible, hence not optimized in terms of execution time). 

A brief flow chart of the program is the following.

\vspace{0.5cm}
\dirtree{%
.1 HOMISBOLTZ().
.2 B\_Constant\_Kernel\_Model(M), {\rmfamily CKM with $\theta=0$ in Eq. (\ref{generic})}.
.3 B\_Constant\_Kernel\_Model\_Creator(M).
.4 C\_Constant\_Kernel\_Model(nEi,nEj,$E_{k\pm}/\Delta E$)$=C_{\pm}$, {\rmfamily see Eq. (\ref{Cpmgeneric})}.
.4 EnergyPerms($\bm{v}$), {\rmfamily where $\bm{v}=[i,j,k,l]^T$ (DVM correction)}.
.4 Read\_B(B,i,j,k,l)$=B_{ij}^{kl}$ {\rmfamily (DVM correction)}.
.4 Write\_B(B,i,j,k,l,$B_{ij}^{kl}$) {\rmfamily (DVM correction)}.
.2 B\_Hard\_Sphere\_Model(M), {\rmfamily HSM with $\theta=1$ in Eq. (\ref{generic})}.
.3 B\_Hard\_Sphere\_Model\_Creator(M).
.4 C\_Hard\_Sphere\_Model(nEi,nEj,$E_{k\pm}/\Delta E$)$=C_{\pm}$, {\rmfamily see Eq. (\ref{Cpmgeneric})}.
.4 EnergyPerms($\bm{v}$), {\rmfamily where $\bm{v}=[i,j,k,l]^T$ (DVM correction)}.
.4 Read\_B(B,i,j,k,l)$=B_{ij}^{kl}$ {\rmfamily (DVM correction)}.
.4 Write\_B(B,i,j,k,l,$B_{ij}^{kl}$) {\rmfamily (DVM correction)}.
.2 A\_Hard\_Sphere\_Model(M), {\rmfamily [it may be omitted by Eq. (\ref{e_AvB2})]}.
.2 Equilibrium($f_i$,$E_i$,$\Delta E$)$=\tilde{f}_E$, {\rmfamily see Eqs. (\ref{eq_E_num}, \ref{eq_E_num_cond})}.
.2 Thermodynamics($f_i$,$\Delta E$)=[$\tilde{n}$,$\tilde{e}$,$k_B\,\tilde{T}$].
.2 Nu\_Equilibrium($F$,$\Delta E$,$E_i$,$(\tilde{f}_E)_i$,$A$)$=(\nu_E)_i$, {\rmfamily see Eq. (\ref{nue})}.
.2 Phi($f_i$,$\Delta E$,$p$)$=\tilde{\Phi}_p$, {\rmfamily see Eq. (\ref{momentp2})}.
}\vspace{0.5cm}

Essentially there are two main parts in the \code{HOMISBOLTZ} program: (a) computing the data structure storing the dimensionless frequencies for the considered model, i.e. \code{B(i,j)}, and (b) the main solution loop (fully explicit and based on the forward Euler integration rule). Both parts are described in the following sections.

\section{\label{components}Description of the individual software components}

\subsection{Data structure storing the dimensionless frequencies}

The data structure storing the dimensionless frequencies, i.e. \code{B(i,j)}, is the fundamental data structure of the whole program and it is also the most time-consuming to be computed. Essentially \code{B(i,j)} is the data structure storing the dimensionless frequencies $\tilde{B}_{ij}^{kl}$ (let us suppose that the DVM correction applies) for all the \emph{GAIN} events $(E_k,E_l)\rightarrow(E_i,E_j)$. The dimensionless frequencies $\tilde{A}_{ij}$ for all the \emph{LOSS} events $(E_i,E_j)\rightarrow(E_k,E_l)$ can be computed by Eq. (\ref{e_AvB2b}). The matrix $\tilde{B}_{ij}^{kl}$ is a four dimensional (sparse) matrix and it is not convenient to compute/store it directly. 

Let us introduce a proper labeling for dealing with the sparse matrix $\tilde{B}_{ij}^{kl}$. Let us define $\Lambda_{ij}$ the set formed by all the pairs of natural indices $(k,l)$ such that $E_k+E_l=E_i+E_j$, with $0< E_k< E_M$ and $0< E_l< E_M$. Let us define with $M_{ij}$ the number of elements of the set $\Lambda_{ij}$ and let us identify each pair of indices by $\lambda$, namely if $1\leq \lambda\leq M_{ij}$ then $(k(\lambda),l(\lambda))\in\Lambda_{ij}$. Consequently, Eq. (\ref{e_coll7}) (after the DVM correction) can be reformulated by reducing the number of nested summations, namely
\begin{equation}\label{e_coll_code}
\tilde{Q}_i= 
\tilde{F}\,\Delta{E}\,\sum_{j=1}^M E_j^{1/2}\,\left(\tilde{\Psi}_{ij}
-f_i f_j\,\tilde{A}_{ij}\right),
\end{equation}
\begin{equation}\label{Psi_code}
\tilde{\Psi}_{ij}=\sum_{\lambda=1}^{M_{ij}} f_{k(\lambda)} f_{l(\lambda)}\,\tilde{B}_{ij}^{k(\lambda)l(\lambda)},
\end{equation}
and $\tilde{B}_{ij}^{k(\lambda)l(\lambda)}$ is simply $\tilde{B}_{ij}^{kl}$ for $k=k(\lambda)$ and $l=l(\lambda)$ and it is stored in the data structure \code{B(i,j)}. Hence all the relevant information stored in \code{B(i,j)} can be labeled by $\lambda$.

A brief overview of the data structure \code{B(i,j)} is the following.

\vspace{0.5cm}
\dirtree{%
.1 B(i,j).
.2 B(i,j).howmany$=M_{ij}$.
.2 B(i,j).k($\lambda$)$=k(\lambda)$.
.2 B(i,j).l($\lambda$)$=l(\lambda)$.
.2 B(i,j).nEkm($\lambda$)$=E_{k(\lambda)-}/\Delta E=k(\lambda)-1$.
.2 B(i,j).nEkp($\lambda$)$=E_{k(\lambda)+}/\Delta E=k(\lambda)$.
.2 B(i,j).value($\lambda$)$=\tilde{B}_{ij}^{k(\lambda)l(\lambda)}${\rmfamily by Eqs. (\ref{e_Bo}, \ref{e_coll_corr}).}.
}\vspace{0.5cm}

\subsection{Main loop}

The main loop of the program aims to compute $\tilde{Q}_i=\tilde{N}_i-\tilde{\nu}_i\,f_i$ at a given time step. Recalling Eq. (\ref{e_coll_code}), the operative formulas immediately follow, namely
\begin{equation}\label{N_code}
\tilde{N}_i= 
\tilde{F}\,\Delta{E}^{3/2}\,\sum_{j=1}^M (j-1/2)^{1/2}\,\tilde{\Psi}_{ij},
\end{equation}
\begin{equation}\label{nu_code}
\tilde{\nu}_i= 
\tilde{F}\,\Delta{E}^{3/2}\,\sum_{j=1}^M (j-1/2)^{1/2}\,f_j\,\tilde{A}_{ij}.
\end{equation}
The previous formulas are implemented straightforwardly in the main loop.

\vspace{0.5cm}
\lstset{language=Matlab}
\lstset{frame=shadowbox}
\begin{lstlisting}
...
for t = ... (time)
    for i = 1:M
        N(i)  = 0;
        nu(i) = 0;
        for j = 1:M
            % GAIN term
            Psi(i,j) = 0;
            for m = 1:B(i,j).howmany
                k = B(i,j).k(m);
                l = B(i,j).l(m);
                Bijkm = B(i,j).value(m);
                Psi(i,j) = Psi(i,j)+f(k)*f(l)*Bijkm;
            end
            N(i)  = N(i)+...
              F*DeltaE^(3/2)*(j-1/2)^(1/2)*Psi(i,j);
            
            % LOSS term
            nu(i) = nu(i)+...
              F*DeltaE^(3/2)*(j-1/2)^(1/2)*f(j)*A(i,j);
        end
        Q(i) = N(i)-nu(i)*f(i);
    end
    % Forward Euler integration rule (explicit)
    f = f+Deltat.*Q;
end
...
\end{lstlisting}

\section{\label{installation}Installation instructions}

The package of the \code{HOMISBOLTZ} program consists of five files, namely
\begin{enumerate}
\item \code{HOMISBOLTZ.m}, which is the (single-file) main program (including all the subroutines described in the previous flow chart);
\item \code{CKM\_Structure\_B\_DVM\_nodes\_50.mat}, which is the binary data file containing the data structure \code{B(i,j)} in case of the Constant Kernel Model (CKM, $\theta=0$ in Eq. (\ref{generic})) with $M=50$;
\item \code{CKM\_Structure\_B\_DVM\_nodes\_100.mat}, which is the binary data file containing the data structure \code{B(i,j)} in case of the Constant Kernel Model (CKM, $\theta=0$ in Eq. (\ref{generic})) with $M=100$;
\item \code{HSM\_Structure\_B\_DVM\_nodes\_50.mat}, which is the binary data file containing the data structure \code{B(i,j)} in case of the Hard Sphere Model (HSM, $\theta=1$ in Eq. (\ref{generic})) with $M=50$;
\item \code{HSM\_Structure\_B\_DVM\_nodes\_100.mat}, which is the binary data file containing the data structure \code{B(i,j)} in case of the Hard Sphere Model (HSM, $\theta=1$ in Eq. (\ref{generic})) with $M=100$.
\end{enumerate}

The previous \code{*.mat} files are not strictly required. When executed, the program first searches for the binary data file corresponding to the required combination of collision kernel, discretization resolution ($M$) and DVM correction (ON/OFF). If this binary data file exists, it will be loaded for saving computational time. Otherwise the data structure \code{B(i,j)} will be computed and saved as binary data file for future use. Hence the previous binary data files are provided as examples.

\section{\label{test}Test run description}

In this section, a full test case is described. Let us consider a dilute gas made of molecules. The interactions among the molecules can be described by means of the collision kernel given by Eq. (\ref{generic}) with the following parameters\footnote{International System of Units (SI) applies. Clearly molecules of $1$ $m$ are not realistic but this dimension was adopted for simplicity. It is important to point out that the characteristic time scale of the relaxation phenomenon $\tau$ scales as $\tau\sim (n\,a^2\,c_s)^{-1} \sim a^{-2}$.}
\begin{equation}\label{parameters}
a=1,\qquad
c_s=50,\qquad
\theta=1\qquad\mbox{Hard Sphere Model (HSM)}.
\end{equation}
No BGK-like approximation is adopted, since we want to investigate the full nonlinear Boltzmann equation (in the homogeneous isotropic case).

From the numerical point of view, we can not investigate the full space $\mathbb{R}^+$ for the particle kinetic energy. We need to bound our investigations in the range $[0,\,E_M]$ and to be sure that the initial conditions well fit into this sub-portion of $\mathbb{R}^+$. Actually, in order to achieve better accuracies, as already pointed out (see Section \ref{DVM} for details), the whole dynamic phenomenon should fit into the sub-portion $[0,E_M/2]$. Let us consider the following initial condition $f(t=0,E)=f_I(E)$, where
\begin{equation}\label{f_0}
f_I(f_0,G_0,G_{00})=f_0\,\exp{\left[-\frac{\left(\sqrt{E}-\sqrt{G_0}\right)^2}{G_{00}}\right]}.
\end{equation}
In the considered test case, the values of these parameters are
\begin{equation}\label{parameters2}
E_M=5000,\qquad
f_0=5\times 10^{-4},\qquad
G_0=600,\qquad
G_{00}=35.
\end{equation}
In order to decide the duration of the numerical simulation, we need to investigate the characteristic time scale of the relaxation phenomenon $\tau$. This characteristic time scales as $\tau\sim (n\,F)^{-1}\sim (n\,a^2\,c_s)^{-1}$. However it may be much smaller than that, when the distribution function approaches the local equilibrium (fluid dynamic regime). Hence the duration of the phenomenon depends also on how far the initial conditions are from the local equilibrium. For the present test case (by trials and errors), the duration of the numerical simulation was fixed at $T_F=1\times 10^{-3}$.

Finally, the parameters concerning the numerical integration must be specified. The range $[0,\,E_M]$ is divided by $M=100$ parts and consequently $\Delta E=E_M/M$. The time frame $T_F$ is divided by $T=100$ parts\footnote{Temperature is not used directly in the code and this ensures that there is no possibility of confusion in the adopted notation.} and consequently $\Delta T=T_F/T$. Since an explicit integration rule is used to solve the kinetic equation (namely the forward Euler rule), an upper threshold on the discretization time step is expected, namely
\begin{equation}\label{CFL}
\Delta T<k_\gamma\,\Delta E^\gamma,
\end{equation}
where $k_\gamma$ is a proper constant and $\gamma$ is an exponent depending on the mode driving the instability ($\gamma=1$ for the advective mode and $\gamma=2$ for the diffusive mode). The previous condition is the celebrated Courant-Friedrichs-Lewy (CFL) stability condition. The adopted parameters for the considered test case satisfy this condition. In the numerical simulations, few non-conserved moments are monitored during the relaxation phenomenon, namely $\tilde{\Phi}_p$ with $p\in[2-9]$ (see Eq. (\ref{momentp2}) for details).

\begin{figure}
\centering
	\includegraphics[width=0.8\textwidth]{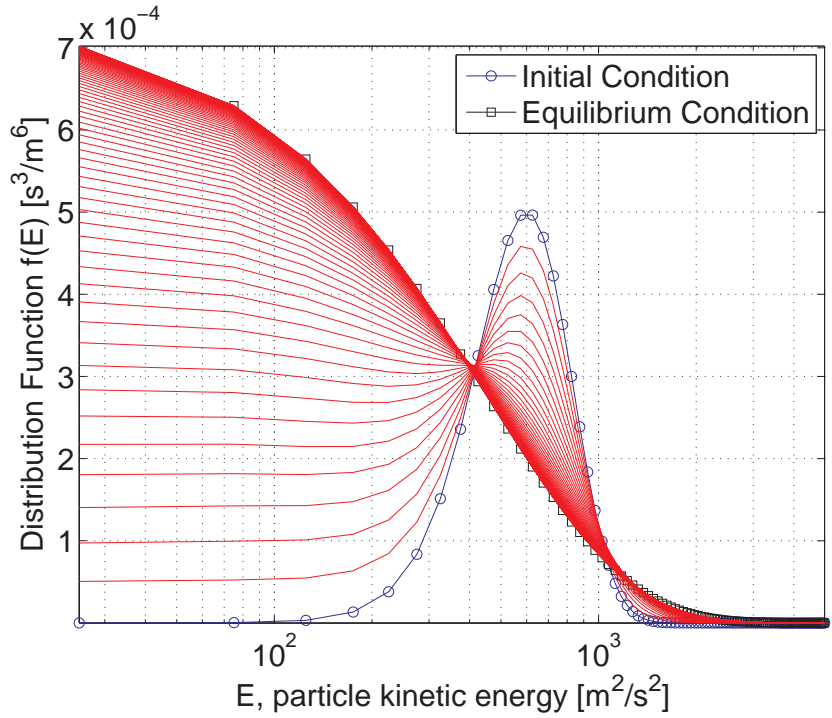}
\caption{(Color online) Distribution function dynamics from the initial condition (blue), namely $f(t=0,E)=f_I(E)$ where $f_I(E)$ is given by Eq. (\ref{f_0}), to the local equilibrium (black), namely $\tilde{f}_E$ given by Eq. (\ref{eq_E_num}, \ref{eq_E_num_cond}). } \label{fig:1}
\end{figure}

Figure \ref{fig:1} reports the distribution function dynamics from the initial condition given by Eq. (\ref{f_0}), to the local equilibrium given by Eq. (\ref{eq_E_num}, \ref{eq_E_num_cond}). The approach to the local equilibrium is initially quite rapid (kinetic stage) and it becomes very slow closer to the equilibrium (fluid dynamic stage). It is not so difficult to catch the main trend in the dynamics of the distribution function. However the formulation in terms of the distribution function may hide some accuracy problems in the relaxation of the high-order moments close to the equilibrium.

\begin{figure}
\centering
	\includegraphics[width=0.8\textwidth]{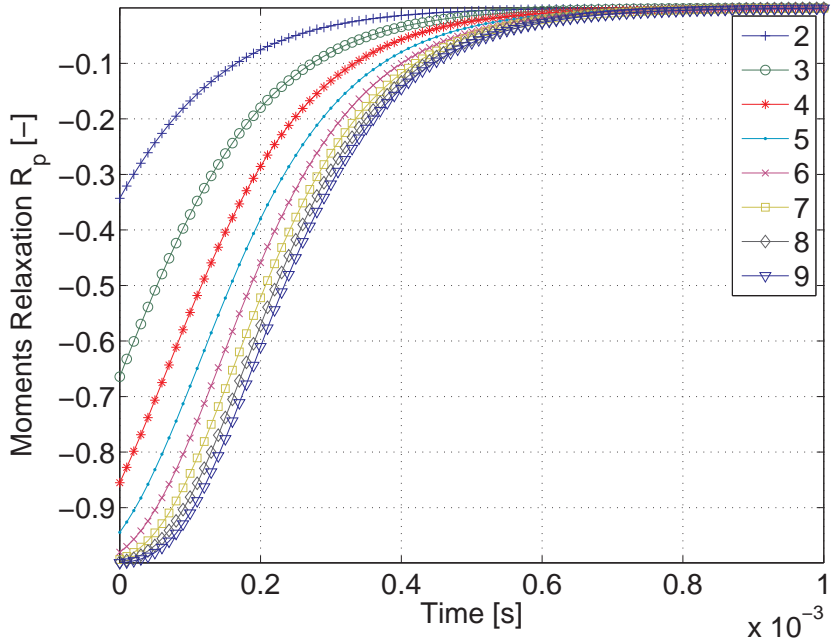}
\caption{(Color online) Macroscopic moments dynamics (in time) described by means of the relaxation rates $\tilde{R}_p$ given by Eq. (\ref{RP}) for $p\in[2-9]$. } \label{fig:2}
\end{figure}

\begin{figure}
\centering
	\includegraphics[width=0.8\textwidth]{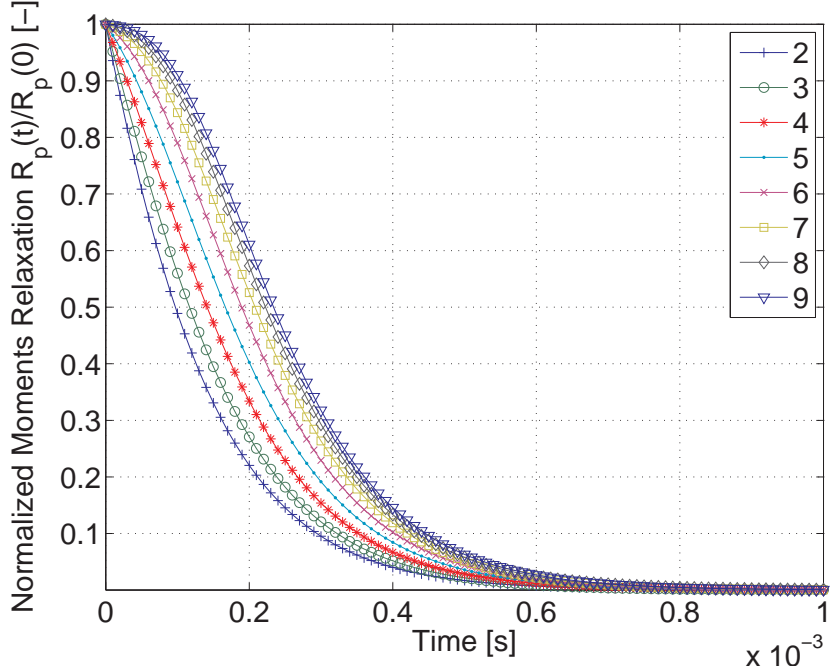}
\caption{(Color online) Normalized macroscopic moments dynamics (in time) described by means of the normalized relaxation rates $\tilde{R}_p/\tilde{R}_p(t=0)$, where $\tilde{R}_p$ is given by Eq. (\ref{RP}) for $p\in[2-9]$. } \label{fig:3}
\end{figure}

In order to investigate the last point, let us introduce the relaxation rate $\tilde{R}_p$ for the macroscopic moment $\tilde{\Phi}_p$, namely
\begin{equation}\label{RP}
\tilde{R}_p=\frac{\tilde{\Phi}_p-\tilde{\Phi}_p^E}{\tilde{\Phi}_p^E},
\end{equation}
where $\tilde{\Phi}_p^E=\tilde{\Phi}_p(\tilde{f}_E)$. The time evolution of the macroscopic moments with $p\in[2-9]$ is described in Figure \ref{fig:2} by means of $\tilde{R}_p$ and in Figure \ref{fig:3} by means of the normalized relaxation rates $\tilde{R}_p/\tilde{R}_p(t=0)$. Both quantities approach the zero value in the late dynamics. The proposed method (and in particular the DVM correction and the recursive tuning of the local equilibrium, see Section \ref{quad} for details) allows one to catch very precisely the approach to the local equilibrium, even by high-order moments. According to the reported results, the hard sphere model produces a slower approach to the equilibrium by the higher order moments. This point is investigated next.

\begin{figure}
\centering
	\includegraphics[width=0.8\textwidth]{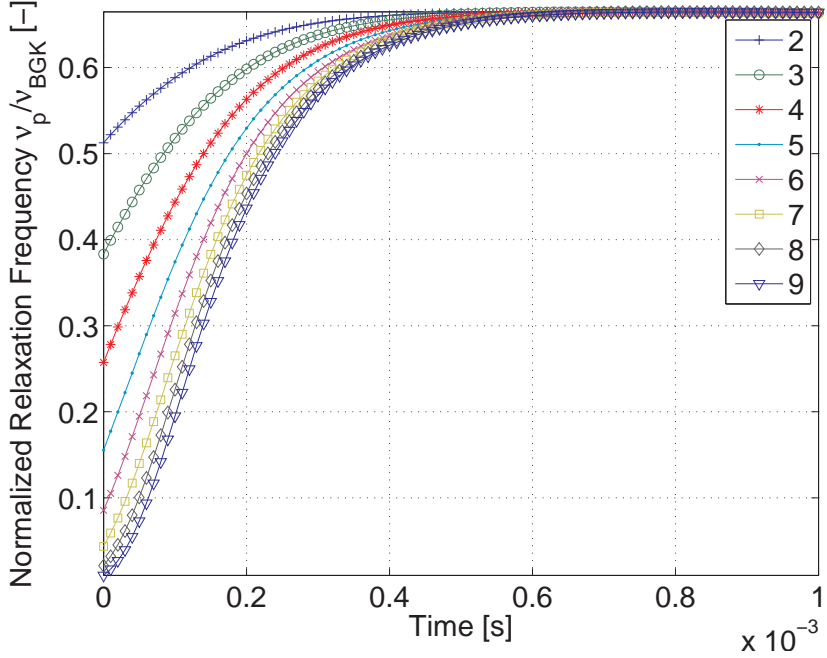}
\caption{(Color online) Time evolution of the normalized effective relaxation frequencies $\tilde{\nu}_p/\nu_{BGK}$ for $p\in[2-9]$, where $\tilde{\nu}_p$ is given by Eq. (\ref{nup}) and $\nu_{BGK}$ is the constant frequency prescribed by the BGK model (see Section \ref{sBGK} for details). } \label{fig:4}
\end{figure}

In order to check even more precisely the late dynamics of the high-order moments, let us introduce a (time-dependent) effective\footnote{Clearly the definition given by Eq. (\ref{nup}) leads to an indeterminate form ($0/0$) for $t\rightarrow\infty$. From the numerical point of view, this may produce some spurious results, particularly when the BGK approximation is used. However, this happens when the quantity $\tilde{\nu}_p$ is no more actually relevant.} relaxation frequency $\tilde{\nu}_p$ for the moment $p$, namely
\begin{equation}\label{nup}
\tilde{\nu}_p=\frac{\tilde{\Phi}_p^Q}{\tilde{\Phi}_p^E-\tilde{\Phi}_p},
\end{equation}
where $\tilde{\Phi}_p^Q=\tilde{\Phi}_p(\tilde{Q})$. In Figure \ref{fig:4} the effective relaxation frequencies for $p\in[2-9]$ are normalized by $\nu_{BGK}$, where $\nu_{BGK}=(\nu_E)_1$, $(\nu_E)_1$ stands for $(\nu_E)_i$ at $E_1=\Delta E/2$ and $(\nu_E)_i$ is given by Eq. (\ref{nue}). The results reported in Figure \ref{fig:4} show that $\tilde{\nu}_p<\nu_{BGK}$ during the whole dynamics and actually $\tilde{\nu}_p/\nu_{BGK}$ all tend to the same asymptotic value ($\approx0.665$) for $t\rightarrow\infty$. In order to explain such behavior, let us consider the BGK-like approximation given by Eq. (\ref{BGKlike}), i.e. $\tilde{Q}_i\approx(\tilde{Q}_{B})_i$, and let us introduce it in the definition of $\tilde{\nu}_p$ (the subscript $i$ has been removed for simplicity), namely
\begin{equation}\label{nup2}
\tilde{\nu}_p\approx\frac{\tilde{\Phi}_p\left(\nu_E\,(\tilde{f}_E-f)\right)}{\tilde{\Phi}_p\left(\tilde{f}_E-f\right)}.
\end{equation}
The previous approximation allows one to interpret $\tilde{\nu}_p$ as a weighted average of the relaxation frequency $(\nu_E)_i$ (valid in the late dynamics) by means of the weight $(\tilde{f}_E-f)_i$. The weight $(\tilde{f}_E-f)_i$ has no definite sign: the ranges of $E_i$ where this weight is positive or negative depend on the initial condition (both ranges must exist because $(\tilde{f}_E)_i$ and $f$ have the same number density by definition). In particular, the adopted initial condition given by Eq. (\ref{f_0}) implies $(\tilde{f}_E-f)_i<0$ for high kinetic energies (see Figure \ref{fig:1}). Taking into account that the relaxation frequency $(\nu_E)_i$ of the hard sphere model for high kinetic energies tends to increase monotonically as $(\nu_E)_i\sim\sqrt{E_i}$ (according to Eq. (\ref{nue_hsm1})), this leads to a penalization effect in the computation of the effective frequency $\tilde{\nu}_p$. This penalization is larger for higher order moments (i.e. it increases with $p$, as showed in Figure \ref{fig:4}) and this explains why the hard sphere model produces a slower approach to the equilibrium by the higher order moments.

\section{Conclusions}

In this work, some improvements to the deterministic numerical method proposed by Aristov \cite{aristov2001} for the homogeneous isotropic Boltzmann equation are discussed. Firstly, the original problem was reformulated in terms of particle kinetic energy and this allows one to ensure exact particle number and energy conservation during the microscopic collisions (momentum is trivially conserved because of the isotropic symmetry). Secondly, the computation of the relaxation rates was improved by the DVM correction, which allows one to satisfy exactly the macroscopic conservation laws and it is particularly suitable for dealing with the late dynamics of the relaxation towards the equilibrium.

This work aims also to distribute an open-source program (called \code{HOMISBOLTZ}), which can be easily understood and modified for dealing with different applications (thermodynamics, econophysics and sociodynamics), in order to derive reliable reference solutions (with an accuracy which can not be easily obtained by stochastic methods). The \code{HOMISBOLTZ} program was developed using Matlab\textregistered. The basic idea is to provide a simple illustration of the discussed methodology, which can be easily ported to other environments (FORTRAN, C++,...). The \code{HOMISBOLTZ} program is free software, which can be redistributed and/or modified under the terms of the GNU General Public License. The \code{HOMISBOLTZ} program has been purposely designed in order to be minimal, not only with regards to the reduced number of lines (less than 1,000), but also with regards to the coding style (as simple as possible, hence not optimized in terms of execution time). 

\section*{Acknowledgements}

The author would like to thank Professor Taku Ohwada (Kyoto University, Japan) for many enlightening clarifications about the solution of the Boltzmann equation by deterministic numerical methods. Moreover he would like to thank Dr. Miguel Onorato and Davide Proment (Universit\`a degli Studi di Torino, Physics Department, Italy) for useful comments. The author acknowledges the support of the EnerGRID project.



\bibliographystyle{elsarticle-num}

\end{document}